\documentclass[aps,prl,nofootinbib]{revtex4}
\usepackage{amsmath}
\usepackage{hyperref}

\begin{document}

\title{Trigonometric identities inspired by atomic form factor}
\author{Abhijit Sen}
\email{abhijit913@gmail.com}
\affiliation{Novosibirsk State University, Novosibirsk 630
090, Russia}

\author{Zurab K.~Silagadze}
\email{Z.K.Silagadze@inp.nsk.su}
\affiliation{ Budker Institute of
Nuclear Physics and Novosibirsk State University, Novosibirsk 630
090, Russia }

\begin{abstract}
We prove some trigonometric identities involving Chebyshev polynomials of
second kind. The identities were inspired by atomic form factor calculations. 
Generalizations of these identities, if found, will help to increase the 
numerical stability of atomic form factors calculations for highly excited 
states.
\end{abstract}

\maketitle

\section{Introduction}
The knowledge of discrete-discrete atomic form factors is important in 
computations of transition probabilities between two different atomic states
when a hydro\-gen-like elementary atom (for example, $\pi^+\pi^-$ or  $\mu^+
\mu^-$) collides with the atom of target material \cite{1A}. A comprehensive 
review of atomic form factors calculations can be found in \cite{1B}.
 
Recently it became evident that the production and study of the never before 
observed true muonium (dimuonium) atom is possible at modern electron-positron 
colliders \cite{1C1,1C2,1C3,1C}, in fixed-target experiments 
\cite{1D1,1D2,1D3,1D4,1D}, in relativistic heavy ion collisions \cite{HI1,HI2},
in a quark-gluon plasma \cite{QG}, in elementary particle decays \cite {D1,D2,
D3,D4,D5,D6}, or using ultra-slow muon beams \cite{USM1,USM2}. Observation of
true muonium signal from astrophysical sources is also of considarable 
interest \cite{ASTR}. 

As a by-product of atomic form factor calculations, which were initiated by 
a modern experimental proposal \cite{1E} in this field, some interesting 
trigonometric identities emerged which we report in this short note.

\section{Integral related to the atomic form factor}
The following integral
\begin{equation}
I_n^m=\int\limits_0^\infty e^{-x}\sin{(\sigma x)}x^m\left[L_n^m(x)\right]^2
dx
\label{eq1}
\end{equation}
arises naturally when calculating certain atomic form factors \cite{1F,1}. Here
\begin{equation}
L_n^m(x)=(n+m)!\sum\limits_{k=0}^n\frac{(-1)^k}{k!(n-k)!(k+m)!}\,x^k
\label{eq2}
\end{equation}
are associated Laguerre polynomials.

In \cite{1} this integral is calculated by using the formula
%\begin{eqnarray} &&
\begin{equation}
\int\limits_0^\infty e^{-bx}x^\alpha\left[L_n^\alpha(\lambda x)
L_m^\alpha(\mu  x)\right]dx= \frac{\Gamma(m+n+\alpha+1)}{\Gamma(m+1)
\Gamma(n+1)}\,\frac{(b-\lambda)^n (b-\mu)^m}{b^{n+m+\alpha+1}}
%\times \nonumber \\ && 
{_2{F}_1}
\left(-m,-n;-m-n-\alpha;\frac{b(b-\lambda-\mu)}{(b-\mu)(b-\lambda)}\right),
\label{eq3}
\end{equation}
%\end{eqnarray}
which can be found in \cite{2}, entry 7.414.4.

The calculation goes as follows. From (\ref{eq3}) we get
\begin{equation}
I_n^m=\frac{(2n+m)!}{(n!)^2}\;\mathrm{Im}\left[\frac{(\mu-1)^{2n}}
{\mu^{2n+m+1}}\,{_2{F}_1}
\left(-n,-n;-2n-m;\frac{\mu(\mu-2)}{(\mu-1)^2}\right)\right],
\label{eq4}
\end{equation}
where $\mu=1-i\sigma$. Let us introduce an angle $\phi$ such that
$\sigma=\tan{\phi}$. Then
\begin{equation}
\frac{\mu(\mu-2)}{(\mu-1)^2}=1+\frac{1}{\sigma^2}=\frac{1}{\sin^2{\phi}},
\label{eq5}
\end{equation}
and
\begin{equation}
\mathrm{Im}\left[\frac{(\mu-1)^{2n}}{\mu^{2n+m+1}}\right ]=
(-1)^n\sin^{2n}{\phi}\cos^{m+1}{\phi}\sin{[(2n+m+1)\phi]},
\label{eq6}
\end{equation}
because $\mu=1-i\sigma=e^{-i\phi}/\cos{\phi}$. Besides we use the following 
identity of the hypergeometric function $_2{F}_1$:
\begin{equation}
{_2{F}_1}(-n,b;c;z)=\frac{(b)_n}{(c)_n}\,(-z)^n\,{_2{F}_1}\left 
(-n,1-c-n;1-b-n;\frac{1}{z}\right ),
\label{eq7}
\end{equation}
where $(x)_n=\Gamma(x+n)/\Gamma(x)$ is the Pochhammer's symbol 
(shifted factorial) with the property $$(-x)_n=(-1)^n(x-n+1)_n.$$ As a result,
equation (\ref{eq4}) can be rewritten in the form
\begin{equation}
\hspace*{-2mm}
I_n^m=\frac{(n+m)!}{n!}\,\cos^{m+1}{\phi}\,\sin{[(2n+m+1)\phi]}\,
{_2F_1}\left(-n,n+m+1;1;\sin^2{\phi}\right).
\label{eq8}
\end{equation}
This result can be expressed in terms of Jacobi polynomials, defined through
the relation
\begin{equation}
P_n^{(\alpha,\beta)}(x)=\frac{(\alpha+1)_n}{n!}\;{_2F_1}\left(-n,n+\alpha+
\beta+1;1+\alpha;\frac{1-x}{2}\right),
\label{eq9}
\end{equation}
and finally we obtain
\begin{equation}
I_n^m=\frac{(n+m)!}{n!}\,\cos^{m+1}{\phi}\,\sin{[(2n+m+1)\phi]}\,
P_n^{(0,m)}(\cos{2\phi}).
\label{eq10}
\end{equation}

\section{Trigonometric identities}
However, we can calculate the integral (\ref{eq1}) in a different way. By 
using Howell's identity \cite{3,4}
\begin{equation}
[L_n^m(x)]^2=\frac{(n+m)!}{2^{2n}n!}\sum\limits_{k=0}^n\frac{(2k)![2(n-k)]!}
{k![(n-k)!]^2(m+k)!}\,L_{2k}^{2m}(2x),
\label{eq11}
\end{equation}
we get
\begin{equation}
I_n^m=\frac{(n+m)!}{2^{2n}n!}\sum\limits_{k=0}^n\frac{(2k)![2(n-k)]!}
{k![(n-k)!]^2(m+k)!}\,J_{m,k},
\label{eq12}
\end{equation}
where 
\begin{equation}
J_{m,k}=\sqrt{\frac{\pi\sigma}{2}}\int\limits_0^\infty e^{-x}J_{1/2}
(\sigma x)x^{m+1/2}L_{2k}^{2m}(2x)dx.
\label{eq13}
\end{equation}
Here $J_\nu(x)$ is the Bessel function of the first kind and we have used
\begin{equation}
J_{1/2}(x)=\sqrt{\frac{2}{\pi x}}\,\sin{x}
\label{eq14}
\end{equation}
to express $\sin{(\sigma x)}$ in terms of the Bessel function.

Integrals of the type (\ref{eq13}) were evaluated in \cite{5} with the result
\begin{eqnarray} && 
\int\limits_0^\infty e^{-\sigma x}J_\nu(\mu x)x^\gamma L_n^\alpha(\beta x)dx=
%\nonumber \\ &&
\sum\limits_{k=0}^n\frac{(-\beta)^k\mu^\nu\Gamma(n+\alpha+1)\Gamma(\nu+
\gamma+k+1)}{k!\Gamma(n-k+1)\Gamma(\alpha+k+1)2^\nu\Gamma(\nu+1)
\sigma^{\nu+\gamma+k+1}} 
\times \nonumber \\ &&
{_2F_1}\left(\frac{\nu+\gamma+k+1}{2},\frac{\nu+
\gamma+k+2}{2};1+\nu;-\frac{\mu^2}{\sigma^2}\right).
\label{eq15}
\end{eqnarray}
By applying this general result to our integral (\ref{eq13}), we get
\begin{equation}
J_{m,k}=\cos^{m+1}{\phi}\sum\limits_{r=0}^{2k}\frac{(-2)^r[2(k+m)]!(m+r)!}
{r!(2k-r)!(2m+r)!}\,\cos^r{\phi}\,\sin{[(m+r+1)\phi]},
\label{eq16}
\end{equation}
where we have taken into account that
\begin{equation}
{_2F_1}\left(a,a+\frac{1}{2};\frac{3}{2};-\tan^2{\phi}\right)=
\cos^{2a}{\phi}\,\frac{\sin{(2a-1)\phi}}{(2a-1)\sin{\phi}}.
\label{eq17}
\end{equation}
Equations (\ref{eq12}) and (\ref{eq16}) imply
%\begin{eqnarray} && \hspace*{-10mm}
\begin{equation} 
I_n^m=\frac{(n+m)!}{2^{2n}n!}\cos^{m+1}{\phi} 
\sum\limits_{k=0}^n\sum\limits_{r=0}^{2k}\frac{(-2)^r(2k)![2(n-k)]!
[2(k+m)]!(m+r)!}{k!r![(n-k)!]^2(m+k)!(2k-r)!(2m+r)!}
%\times \nonumber \\ && 
\cos^r{\phi}\,\sin{[(m+r+1)\phi]}. 
\label{eq18}
\end{equation}
%\end{eqnarray}
Comparing this result with (\ref{eq10}), we get the following trigonometric 
identity
%\begin{eqnarray} && 
\begin{equation}
\sum\limits_{k=0}^n\sum\limits_{r=0}^{2k}(-2)^r
\frac{(k+1)_m}{(m+r+1)_m}\binom{2(n-k)}{n-k}\binom{2k}{r}
\binom{2(m+k)}{m+k}
%\times\nonumber \\ &&
\cos^r{\phi}\,U_{m+r}(\cos{\phi})=
2^{2n}U_{2n+m}(\cos{\phi})\,P_n^{(0,m)}(\cos{2\phi}),
\label{eq19}
\end{equation}
%\end{eqnarray}
where 
\begin{equation}
U_n(\cos{\phi})=\frac{\sin{[(n+1)\phi]}}{\sin{\phi}}
\label{eq21}
\end{equation}
are Chebyshev polynomials of the second kind.

In particular, when $m=0$ we get
%\begin{eqnarray} &&
\begin{equation}
\sum\limits_{k=0}^n\sum\limits_{r=0}^{2k}(-2)^r
\binom{2(n-k)}{n-k}\binom{2k}{r}
\binom{2k}{k}\cos^r{\phi}\,U_r(\cos{\phi})= 
%\nonumber \\ &&
2^{2n}\,U_{2n}(\cos{\phi})\,P_n(\cos{2\phi}).
\label{eq20}
\end{equation}
%\end{eqnarray}
Here $P_n(x)=P_n^{(0,0)}(x)$ are Legendre polynomials.

\section{Further trigonometric identities}
One more possibility to calculate (\ref{eq1}) and obtain other trigonometric 
identities is to expand one of the Laguerre polynomials in (\ref{eq1})
according to (\ref{eq2}) and then use (\ref{eq15}). In this way we get
%\begin{eqnarray} &&
\begin{equation}
I_n^m=\left[\frac{(n+m)!}{n!}\right]^2\cos^{m+1}{\phi}
\sum\limits_{k=0}^n\sum\limits_{r=0}^n\frac{(-1)^{k+r}(k+r+1)_m}
{(k+1)_m(r+1)_m}
%\times\nonumber \\ && 
\binom{n}{k}\binom{n}{r}\binom{k+r}{k}
\cos^{k+r}{\phi}\,\sin{[(m+k+r+1)\phi]}.
\label{eq22}
\end{equation}
%\end{eqnarray}
In light of (\ref{eq10}), this result implies the validity of the following
trigonometric identity
%\begin{eqnarray} &&
\begin{equation}
\sum\limits_{k=0}^n\sum\limits_{r=0}^n
\frac{(-1)^{k+r}(n+1)_m(k+r+1)_m}{(k+1)_m(r+1)_m}
\binom{n}{k}\binom{n}{r}\binom{k+r}{k}
%\times\nonumber \\ &&
\cos^{k+r}{\phi}\,U_{k+r+m}(\cos{\phi})=
U_{2n+m}(\cos{\phi})\,P_n^{(0,m)}(\cos{2\phi}).
\label{eq23}
\end{equation}
%\end{eqnarray}
Its special case when $m=0$ can be written as follows
%\begin{eqnarray} &&
\begin{equation}
\sum\limits_{k=0}^n\sum\limits_{r=0}^n
(-1)^{k+r}\binom{n}{k}\binom{n}{r}\binom{k+r}{k}\cos^{k+r}{\phi}\;U_{k+r}(
\cos{\phi})= 
%\nonumber \\ &&
U_{2n}(\cos{\phi})\,P_n(\cos{2\phi}).
\label{eq24}
\end{equation}
%\end{eqnarray}

Feldheim identities \cite{5A,5B} provide still another possibility 
to generate further trigonometric identities. According to these identities
\begin{equation}
[L_n^m(x)]^2=\sum\limits_{k=0}^{2n}\sum\limits_{r=0}^{k}(-1)^k
\binom{k}{r}\binom{n+m}{n-k+r}\binom{n+m}{n-r}L_k^{2m}(x),
\label{eq11A}
\end{equation}
and
\begin{equation}
[L_n^m(x)]^2=\sum\limits_{k=0}^{2n}\sum\limits_{r=0}^{k}(-1)^k
\binom{k}{r}\binom{n+m}{n-k+r}\binom{n+m}{n-r}\frac{x^k}{k!}.
\label{eq11B}
\end{equation}
The first expansion in a similar way as described above leads to the identity
%\begin{eqnarray} &&
\begin{equation}
\sum\limits_{k=0}^{2n}\sum\limits_{r=0}^k\sum\limits_{s=0}^k
(-1)^{k+s}\frac{(n+1)_m}{(m+s+1)_m}
\binom{n}{r}\binom{n}{k-r}\binom{k}{s}\binom{2m+k}{m+r}
%\times\nonumber \\ &&
\cos^s{\phi}\,U_{m+s}(\cos{\phi})= U_{2n+m}(\cos{\phi})\,P_n^{(0,m)}
(\cos{2\phi}).
\label{eq23A}
\end{equation}
%\end{eqnarray}
The second identity (\ref{eq11B}) can be used in conjnucation to the integral
(which follows from the entry 6.621.1 in \cite{2})
\begin{equation}
\int_0^\infty e^{-x}J_{1/2}(\sigma x)x^{m+k+1/2}=\sqrt{\frac{\sigma}{2}}\,
\frac{(m+k+1)!}{\Gamma(3/2)}\,{_2F_1}\left(a,a+\frac{1}{2};\frac{3}{2};
-\sigma^2\right),
\label{eq25A}
\end{equation}
where $a=(m+k+2)/2$. As a result, we end up with
%\begin{eqnarray} &&
\begin{equation}
\sum\limits_{k=0}^{2n}\sum\limits_{r=0}^k
(-1)^k\frac{(k+1)_m}{(n+1)_m}
\binom{k}{r}\binom{n+m}{n-k+r}\binom{n+m}{n-r}
%\times\nonumber \\ &&
\cos^k{\phi}\,U_{m+k}(\cos{\phi})=
U_{2n+m}(\cos{\phi})\,P_n^{(0,m)}(\cos{2\phi}).
\label{eq23B}
\end{equation}
%\end{eqnarray}
The $m=0$ special cases of these new identities are
%\begin{eqnarray} &&
\begin{equation}
\sum\limits_{k=0}^{2n}\sum\limits_{r=0}^k\sum\limits_{s=0}^k
(-1)^{k+s}\binom{n}{r}\binom{n}{k-r}\binom{k}{s}\binom{k}{r}
%\times\nonumber \\ && 
\cos^s{\phi}\,U_s(\cos{\phi})=
U_{2n}(\cos{\phi})\,P_n(\cos{2\phi}),
\label{eq24A}
\end{equation}
%\end{eqnarray}
and
%\begin{eqnarray} &&
\begin{equation}
\sum\limits_{k=0}^{2n}\sum\limits_{r=0}^k
(-1)^k\binom{k}{r}\binom{n}{n-k+r}\binom{n}{n-r}\cos^k{\phi}\,U_k(\cos{\phi})=
%\nonumber \\ &&
U_{2n}(\cos{\phi})\,P_n(\cos{2\phi}).
\label{eq24B}
\end{equation}
%\end{eqnarray}

\section{Concluding remarks}
Trigonometric identities (\ref{eq19}), (\ref{eq23}), (\ref{eq23A}) and 
(\ref{eq23B}), as well as their particular cases (\ref{eq20}), (\ref{eq24}),
(\ref{eq24A}) and (\ref{eq24B}), are the main results of this 
work. These identities are equivalent to the statement that the integral
(\ref{eq1}) can be calculated in the compact form  (\ref{eq10}) which is 
very convenient for numerical evaluation, because Jacobi polynomials can
be calculated recursively. 

While calculating more general atomic form factors, the integral (\ref{eq1})
generalizes to
\begin{equation}
I_n^{m,k}=\sigma\int\limits_0^\infty e^{-x}j_k(\sigma x)x^{m+1-k}
\left[L_n^m(x)\right]^2 dx,
\label{eq25}
\end{equation}
where $j_k(x)$ is the spherical Bessel function. Note that $j_0(x)=\sin{x}/x$
and correspondingly $I_n^{m,0}=I_n^m$. If we expand one of the Laguerre 
polynomials in this integral according to (\ref{eq2}) and then use 
(\ref{eq15}), we get
\begin{eqnarray} && \hspace*{-5mm}
I_n^{m,k}=\frac{2^kk!\,\sigma^{k+1}}{(2k+1)!}\left[\frac{(n+m)!}{n!}\right]^2
\sum_{m_1=0}^n\;\sum_{m_2=0}^n\frac{(-1)^{m_1+m_2}(m_1+m_2+1)_m}
{(m_1+1)_m(m_2+1)_m}
\times\nonumber\\ && \hspace*{-12mm}
\binom{n}{m_1}\binom{n}{m_2}\binom{m_1+m_2}{m_1}(m+m_1+m_2+1)\,
{_2{F}_1}\left(a,a+\frac{1}{2};k+\frac{3}{2};-\sigma^2\right)\hspace*{-1mm},
\label{eq26}
\end{eqnarray}
where $a=1+(m+m_1+m_2)/2$. In principle, such kind of expressions can be used
in atomic transition form factor evaluation and a general formulas for the
atomic transition form factor along these lines were found in \cite{1F,1,6}.
The calculations were based on the linearization formulas for the product of 
two Laguerre polynomials \cite{8}. The final results were expressed in terms 
of either Jacobi polynomials \cite{1F,1} or generalized Gegenbauer polynomials
\cite{6} and their computation was implemented using the corresponding 
recurrence relations. However, due to numerical problems related to a near 
cancellation of large numbers in alternating sum, in practice this method 
gives reliable results for all form factors only for relatively small 
principal quantum numbers $n\sim 10$ \cite{7}. 

Calculations of the atomic form factors in \cite{1F,1,6} were motivated by
the needs of the DIRAC experiment \cite{9}. Correspondingly the authors of 
\cite{1F,1,6} never had any numerical problems in computation of the specified 
atomic form factors because such computation was required only for the values 
of the principal quantum number $n\le n_{max}\sim 7-10$ \cite{6}. Anyway for 
DIRAC experiment more pressing issue in increasing the accuracy of the 
Monte-Carlo simulation of the passage of dimesoatoms through matter  was not 
inclusion of the highly excited states, but the inclusion of interference 
effects between the different dimesoatomic states during their passage through 
matter \cite{10}.

Nevetheless it will be of practical interest if the results of this note 
can be generalized for sums of the type (\ref{eq26}). Such a generalization, 
if found, will allow to increase numerical stability of atomic  form factors 
calculations for highly excited states for which the direct computation with 
the existing methods might break down \cite{1B}. The trigonometric identities
obtained in this note are of not direct use in the true muonium and other 
elementary atoms studies. However the method from which they originated allows
an alternative numerical algorithm for calculation of atomic form factors
and its realization will be useful in various correctness tests of computer
programs designed for the atomic  form factors calculations.   

\section*{Acknowledgments}
We thank anonymous Referee for constructive remarks.
The work is supported by the Ministry of Education and 
Science of the Russian Federation.

\bibliography{trig}

\begin{thebibliography}{37}
\expandafter\ifx\csname natexlab\endcsname\relax\def\natexlab#1{#1}\fi
\expandafter\ifx\csname bibnamefont\endcsname\relax
  \def\bibnamefont#1{#1}\fi
\expandafter\ifx\csname bibfnamefont\endcsname\relax
  \def\bibfnamefont#1{#1}\fi
\expandafter\ifx\csname citenamefont\endcsname\relax
  \def\citenamefont#1{#1}\fi
\expandafter\ifx\csname url\endcsname\relax
  \def\url#1{\texttt{#1}}\fi
\expandafter\ifx\csname urlprefix\endcsname\relax\def\urlprefix{URL }\fi
\providecommand{\bibinfo}[2]{#2}
\providecommand{\eprint}[2][]{\url{#2}}

\bibitem[{\citenamefont{Mr\'{o}wczy\'{n}ski}(1986)}]{1A}
\bibinfo{author}{\bibfnamefont{S.}~\bibnamefont{Mr\'{o}wczy\'{n}ski}},
  \bibinfo{journal}{Phys. Rev.} \textbf{\bibinfo{volume}{A33}},
  \bibinfo{pages}{1549} (\bibinfo{year}{1986}).

\bibitem[{\citenamefont{Dewangan}(2012)}]{1B}
\bibinfo{author}{\bibfnamefont{D.}~\bibnamefont{Dewangan}},
  \bibinfo{journal}{Phys. Rep.} \textbf{\bibinfo{volume}{511}},
  \bibinfo{pages}{1} (\bibinfo{year}{2012}).

\bibitem[{\citenamefont{Baier and Synakh}(1962)}]{1C1}
\bibinfo{author}{\bibfnamefont{V.}~\bibnamefont{Baier}} \bibnamefont{and}
  \bibinfo{author}{\bibfnamefont{V.}~\bibnamefont{Synakh}},
  \bibinfo{journal}{Soviet Physics JETP} \textbf{\bibinfo{volume}{14}},
  \bibinfo{pages}{1122} (\bibinfo{year}{1962}).

\bibitem[{\citenamefont{Bilenky et~al.}(1969)\citenamefont{Bilenky, Nguyen,
  Nemenov, and Tkebuchava}}]{1C2}
\bibinfo{author}{\bibfnamefont{S.}~\bibnamefont{Bilenky}},
  \bibinfo{author}{\bibfnamefont{V.}~\bibnamefont{Nguyen}},
  \bibinfo{author}{\bibfnamefont{L.}~\bibnamefont{Nemenov}}, \bibnamefont{and}
  \bibinfo{author}{\bibfnamefont{F.}~\bibnamefont{Tkebuchava}},
  \bibinfo{journal}{Yad. Fiz.} \textbf{\bibinfo{volume}{10}},
  \bibinfo{pages}{812} (\bibinfo{year}{1969}).

\bibitem[{\citenamefont{Moffat}(1975)}]{1C3}
\bibinfo{author}{\bibfnamefont{J.}~\bibnamefont{Moffat}},
  \bibinfo{journal}{Phys. Rev. Lett.} \textbf{\bibinfo{volume}{35}},
  \bibinfo{pages}{1605} (\bibinfo{year}{1975}).

\bibitem[{\citenamefont{Brodsky and Lebed}(2009)}]{1C}
\bibinfo{author}{\bibfnamefont{S.}~\bibnamefont{Brodsky}} \bibnamefont{and}
  \bibinfo{author}{\bibfnamefont{R.}~\bibnamefont{Lebed}},
  \bibinfo{journal}{Phys. Rev. Lett.} \textbf{\bibinfo{volume}{102}},
  \bibinfo{pages}{213401} (\bibinfo{year}{2009}).

\bibitem[{\citenamefont{Holvik and Olsen}(2000)}]{1D1}
\bibinfo{author}{\bibfnamefont{E.}~\bibnamefont{Holvik}} \bibnamefont{and}
  \bibinfo{author}{\bibfnamefont{H.}~\bibnamefont{Olsen}},
  \bibinfo{journal}{Phys. Rev. A} \textbf{\bibinfo{volume}{62}},
  \bibinfo{pages}{032501} (\bibinfo{year}{2000}).

\bibitem[{\citenamefont{Arteaga-Romero
  et~al.}(1987)\citenamefont{Arteaga-Romero, Carimalo, and Serbo}}]{1D2}
\bibinfo{author}{\bibfnamefont{N.}~\bibnamefont{Arteaga-Romero}},
  \bibinfo{author}{\bibfnamefont{C.}~\bibnamefont{Carimalo}}, \bibnamefont{and}
  \bibinfo{author}{\bibfnamefont{V.}~\bibnamefont{Serbo}},
  \bibinfo{journal}{Phys. Rev. D} \textbf{\bibinfo{volume}{35}},
  \bibinfo{pages}{2124} (\bibinfo{year}{1987}).

\bibitem[{\citenamefont{Sakimoto}(2015)}]{1D3}
\bibinfo{author}{\bibfnamefont{K.}~\bibnamefont{Sakimoto}},
  \bibinfo{journal}{Eur. Phys. J. D} \textbf{\bibinfo{volume}{69}},
  \bibinfo{pages}{276} (\bibinfo{year}{2015}).

\bibitem[{\citenamefont{Krachkov and Milstein}(2018)}]{1D4}
\bibinfo{author}{\bibfnamefont{P.}~\bibnamefont{Krachkov}} \bibnamefont{and}
  \bibinfo{author}{\bibfnamefont{A.}~\bibnamefont{Milstein}},
  \bibinfo{journal}{Nucl. Phys. A} \textbf{\bibinfo{volume}{971}},
  \bibinfo{pages}{71} (\bibinfo{year}{2018}).

\bibitem[{\citenamefont{Banburski and Schuster}(2012)}]{1D}
\bibinfo{author}{\bibfnamefont{A.}~\bibnamefont{Banburski}} \bibnamefont{and}
  \bibinfo{author}{\bibfnamefont{P.}~\bibnamefont{Schuster}},
  \bibinfo{journal}{Phys. Rev. D} \textbf{\bibinfo{volume}{86}},
  \bibinfo{pages}{093007} (\bibinfo{year}{2012}).

\bibitem[{\citenamefont{Ginzburg et~al.}(1998)\citenamefont{Ginzburg,
  Jentschura, Karshenboim, Krauss, Serbo, and Soff}}]{HI1}
\bibinfo{author}{\bibfnamefont{I.}~\bibnamefont{Ginzburg}},
  \bibinfo{author}{\bibfnamefont{U.}~\bibnamefont{Jentschura}},
  \bibinfo{author}{\bibfnamefont{S.}~\bibnamefont{Karshenboim}},
  \bibinfo{author}{\bibfnamefont{F.}~\bibnamefont{Krauss}},
  \bibinfo{author}{\bibfnamefont{V.}~\bibnamefont{Serbo}}, \bibnamefont{and}
  \bibinfo{author}{\bibfnamefont{G.}~\bibnamefont{Soff}},
  \bibinfo{journal}{Phys. Rev. C} \textbf{\bibinfo{volume}{58}},
  \bibinfo{pages}{3565} (\bibinfo{year}{1998}).

\bibitem[{\citenamefont{Yu and Li}(2013)}]{HI2}
\bibinfo{author}{\bibfnamefont{G.~M.} \bibnamefont{Yu}} \bibnamefont{and}
  \bibinfo{author}{\bibfnamefont{Y.~D.} \bibnamefont{Li}},
  \bibinfo{journal}{Chin. Phys. Lett.} \textbf{\bibinfo{volume}{30}},
  \bibinfo{pages}{011201} (\bibinfo{year}{2013}).

\bibitem[{\citenamefont{Chen and Zhuang}(2012)}]{QG}
\bibinfo{author}{\bibfnamefont{Y.}~\bibnamefont{Chen}} \bibnamefont{and}
  \bibinfo{author}{\bibfnamefont{P.}~\bibnamefont{Zhuang}},
  \bibinfo{journal}{arXiv preprint 1204.4389}  (\bibinfo{year}{2012}),
  \eprint{1204.4389}, \urlprefix\url{https://arxiv.org/abs/1204.4389}.

\bibitem[{\citenamefont{Nemenov}(1972)}]{D1}
\bibinfo{author}{\bibfnamefont{L.}~\bibnamefont{Nemenov}},
  \bibinfo{journal}{Yad. Fiz.} \textbf{\bibinfo{volume}{15}},
  \bibinfo{pages}{1047} (\bibinfo{year}{1972}).

\bibitem[{\citenamefont{Kozlov}(1988)}]{D2}
\bibinfo{author}{\bibfnamefont{G.}~\bibnamefont{Kozlov}},
  \bibinfo{journal}{Sov. J. Nucl. Phys.} \textbf{\bibinfo{volume}{48}},
  \bibinfo{pages}{167} (\bibinfo{year}{1988}).

\bibitem[{\citenamefont{Ji and Lamm}(2018{\natexlab{a}})}]{D3}
\bibinfo{author}{\bibfnamefont{Y.}~\bibnamefont{Ji}} \bibnamefont{and}
  \bibinfo{author}{\bibfnamefont{H.}~\bibnamefont{Lamm}},
  \bibinfo{journal}{Phys. Rev. D} \textbf{\bibinfo{volume}{98}},
  \bibinfo{pages}{053008} (\bibinfo{year}{2018}{\natexlab{a}}).

\bibitem[{\citenamefont{Lamm and Ji}(2018)}]{D4}
\bibinfo{author}{\bibfnamefont{H.}~\bibnamefont{Lamm}} \bibnamefont{and}
  \bibinfo{author}{\bibfnamefont{Y.}~\bibnamefont{Ji}}, \bibinfo{journal}{EPJ
  Web Conf.} \textbf{\bibinfo{volume}{181}}, \bibinfo{pages}{01016}
  (\bibinfo{year}{2018}).

\bibitem[{\citenamefont{Ji and Lamm}(2018{\natexlab{b}})}]{D5}
\bibinfo{author}{\bibfnamefont{Y.}~\bibnamefont{Ji}} \bibnamefont{and}
  \bibinfo{author}{\bibfnamefont{H.}~\bibnamefont{Lamm}},
  \bibinfo{journal}{arXiv preprint 1810.00233}
  (\bibinfo{year}{2018}{\natexlab{b}}), \eprint{1810.00233},
  \urlprefix\url{https://arxiv.org/abs/1810.00233}.

\bibitem[{\citenamefont{Fael and Mannel}(2018)}]{D6}
\bibinfo{author}{\bibfnamefont{M.}~\bibnamefont{Fael}} \bibnamefont{and}
  \bibinfo{author}{\bibfnamefont{T.}~\bibnamefont{Mannel}},
  \bibinfo{journal}{Nucl. Phys. B} \textbf{\bibinfo{volume}{932}},
  \bibinfo{pages}{370} (\bibinfo{year}{2018}).

\bibitem[{\citenamefont{Nagamine}(2014)}]{USM1}
\bibinfo{author}{\bibfnamefont{K.}~\bibnamefont{Nagamine}},
  \bibinfo{journal}{JPS Conf. Proc.} \textbf{\bibinfo{volume}{2}},
  \bibinfo{pages}{010001} (\bibinfo{year}{2014}).

\bibitem[{\citenamefont{Itahashi et~al.}(2015)\citenamefont{Itahashi, Sakamoto,
  Sato, and Takahisa}}]{USM2}
\bibinfo{author}{\bibfnamefont{T.}~\bibnamefont{Itahashi}},
  \bibinfo{author}{\bibfnamefont{H.}~\bibnamefont{Sakamoto}},
  \bibinfo{author}{\bibfnamefont{A.}~\bibnamefont{Sato}}, \bibnamefont{and}
  \bibinfo{author}{\bibfnamefont{K.}~\bibnamefont{Takahisa}},
  \bibinfo{journal}{JPS Conf. Proc.} \textbf{\bibinfo{volume}{8}},
  \bibinfo{pages}{025004} (\bibinfo{year}{2015}).

\bibitem[{\citenamefont{Ellis and Bland-Hawthorn}(2018)}]{ASTR}
\bibinfo{author}{\bibfnamefont{S.}~\bibnamefont{Ellis}} \bibnamefont{and}
  \bibinfo{author}{\bibfnamefont{J.}~\bibnamefont{Bland-Hawthorn}},
  \bibinfo{journal}{Eur. Phys. J. D} \textbf{\bibinfo{volume}{72}},
  \bibinfo{pages}{18} (\bibinfo{year}{2018}).

\bibitem[{\citenamefont{Bogomyagkov et~al.}(2018)\citenamefont{Bogomyagkov,
  Druzhinin, Levichev, Milstein, and Sinyatkin}}]{1E}
\bibinfo{author}{\bibfnamefont{A.}~\bibnamefont{Bogomyagkov}},
  \bibinfo{author}{\bibfnamefont{V.}~\bibnamefont{Druzhinin}},
  \bibinfo{author}{\bibfnamefont{E.}~\bibnamefont{Levichev}},
  \bibinfo{author}{\bibfnamefont{A.}~\bibnamefont{Milstein}}, \bibnamefont{and}
  \bibinfo{author}{\bibfnamefont{S.}~\bibnamefont{Sinyatkin}},
  \bibinfo{journal}{EPJ Web Conf.} \textbf{\bibinfo{volume}{181}},
  \bibinfo{pages}{01032} (\bibinfo{year}{2018}).

\bibitem[{\citenamefont{Afanasyev and Tarasov}(1993)}]{1F}
\bibinfo{author}{\bibfnamefont{L.}~\bibnamefont{Afanasyev}} \bibnamefont{and}
  \bibinfo{author}{\bibfnamefont{A.}~\bibnamefont{Tarasov}},
  \bibinfo{journal}{JINR Preprint E4-93-293}  (\bibinfo{year}{1993}),
  \bibinfo{note}{also in the book (pp. 90-93):},
  \urlprefix\url{http://www1.jinr.ru/Books/book\_Tarasov.pdf}.

\bibitem[{\citenamefont{Afanasyev et~al.}(1999)\citenamefont{Afanasyev,
  Tarasov, and Voskresenskaya}}]{1}
\bibinfo{author}{\bibfnamefont{L.}~\bibnamefont{Afanasyev}},
  \bibinfo{author}{\bibfnamefont{A.}~\bibnamefont{Tarasov}}, \bibnamefont{and}
  \bibinfo{author}{\bibfnamefont{O.}~\bibnamefont{Voskresenskaya}},
  \bibinfo{journal}{J. Phys. G} \textbf{\bibinfo{volume}{25}},
  \bibinfo{pages}{B7} (\bibinfo{year}{1999}).

\bibitem[{\citenamefont{Gradshteyn and Ryzhik}(1965)}]{2}
\bibinfo{author}{\bibfnamefont{I.}~\bibnamefont{Gradshteyn}} \bibnamefont{and}
  \bibinfo{author}{\bibfnamefont{I.}~\bibnamefont{Ryzhik}},
  \emph{\bibinfo{title}{Table of Integrals, Series, and Products}}
  (\bibinfo{publisher}{Academic Press}, \bibinfo{address}{London},
  \bibinfo{year}{1965}).

\bibitem[{\citenamefont{Bailey}(1939)}]{3}
\bibinfo{author}{\bibfnamefont{W.}~\bibnamefont{Bailey}},
  \bibinfo{journal}{Quart. J. Math.} \textbf{\bibinfo{volume}{os-10}},
  \bibinfo{pages}{60} (\bibinfo{year}{1939}).

\bibitem[{\citenamefont{Carlitz}(1961)}]{4}
\bibinfo{author}{\bibfnamefont{L.}~\bibnamefont{Carlitz}}, \bibinfo{journal}{J.
  London Math. Soc.} \textbf{\bibinfo{volume}{s1-36}}, \bibinfo{pages}{399}
  (\bibinfo{year}{1961}).

\bibitem[{\citenamefont{Alassar et~al.}(2008)\citenamefont{Alassar, Mavromatis,
  and Sofianos}}]{5}
\bibinfo{author}{\bibfnamefont{R.}~\bibnamefont{Alassar}},
  \bibinfo{author}{\bibfnamefont{H.}~\bibnamefont{Mavromatis}},
  \bibnamefont{and} \bibinfo{author}{\bibfnamefont{S.}~\bibnamefont{Sofianos}},
  \bibinfo{journal}{Acta Appl. Math.} \textbf{\bibinfo{volume}{100}},
  \bibinfo{pages}{263} (\bibinfo{year}{2008}).

\bibitem[{\citenamefont{Feldheim}(1940)}]{5A}
\bibinfo{author}{\bibfnamefont{E.}~\bibnamefont{Feldheim}},
  \bibinfo{journal}{Quart. J. Math.} \textbf{\bibinfo{volume}{os-11}},
  \bibinfo{pages}{18} (\bibinfo{year}{1940}).

\bibitem[{\citenamefont{Popov and Srivastava}(2003)}]{5B}
\bibinfo{author}{\bibfnamefont{B.}~\bibnamefont{Popov}} \bibnamefont{and}
  \bibinfo{author}{\bibfnamefont{H.}~\bibnamefont{Srivastava}},
  \bibinfo{journal}{Facta Univ. Ser. Math. Inform.}
  \textbf{\bibinfo{volume}{18}}, \bibinfo{pages}{1} (\bibinfo{year}{2003}).

\bibitem[{\citenamefont{Afanasyev and Tarasov}(1996)}]{6}
\bibinfo{author}{\bibfnamefont{L.}~\bibnamefont{Afanasyev}} \bibnamefont{and}
  \bibinfo{author}{\bibfnamefont{A.}~\bibnamefont{Tarasov}},
  \bibinfo{journal}{Phys. At. Nucl.} \textbf{\bibinfo{volume}{59}},
  \bibinfo{pages}{2130} (\bibinfo{year}{1996}).

\bibitem[{\citenamefont{Prudnikov et~al.}(1983)\citenamefont{Prudnikov,
  Brychkov, and O.I}}]{8}
\bibinfo{author}{\bibfnamefont{A.}~\bibnamefont{Prudnikov}},
  \bibinfo{author}{\bibfnamefont{Y.}~\bibnamefont{Brychkov}}, \bibnamefont{and}
  \bibinfo{author}{\bibfnamefont{O.~M.} \bibnamefont{O.I}},
  \emph{\bibinfo{title}{Integrals and Series. Special Functions}}
  (\bibinfo{publisher}{Nauka}, \bibinfo{address}{Moscow},
  \bibinfo{year}{1983}).

\bibitem[{\citenamefont{R\'{i}os and Silva}(2003)}]{7}
\bibinfo{author}{\bibfnamefont{C.~S.} \bibnamefont{R\'{i}os}} \bibnamefont{and}
  \bibinfo{author}{\bibfnamefont{J.~S.} \bibnamefont{Silva}},
  \bibinfo{journal}{Comp. Phys. Commun.} \textbf{\bibinfo{volume}{151}},
  \bibinfo{pages}{79} (\bibinfo{year}{2003}).

\bibitem[{\citenamefont{Afanasyev}(2016)}]{9}
\bibinfo{author}{\bibfnamefont{L.}~\bibnamefont{Afanasyev}},
  \bibinfo{journal}{Nucl. Part. Phys. Proc.}
  \textbf{\bibinfo{volume}{273-275}}, \bibinfo{pages}{1997}
  (\bibinfo{year}{2016}).

\bibitem[{\citenamefont{Voskresenskaya}(2003)}]{10}
\bibinfo{author}{\bibfnamefont{O.}~\bibnamefont{Voskresenskaya}},
  \bibinfo{journal}{J. Phys. B: At. Mol. Opt. Phys.}
  \textbf{\bibinfo{volume}{36}}, \bibinfo{pages}{3293} (\bibinfo{year}{2003}).

\end{thebibliography}

\end{document}